\documentclass[osajnl,twocolumn,showpacs,superscriptaddress,10pt]{revtex4-1} 
\usepackage{amsmath,amssymb,graphicx}
\usepackage{epstopdf}
\usepackage{color}

\renewcommand{\section}[1]{\textit{#1} --}

\begin{document}

\title{Engineering wavefront caustics trajectories in ${\cal PT}$-symmetric lattices}

\author{Nicholas Bender}
\affiliation{Department of Physics, Wesleyan University, Middletown CT-06457, USA}
\author{Hamidreza Ramezani}
\affiliation{NSF Nanoscale Science and Engineering Center, University of California, Berkeley, California 94720, USA}
\author{Tsampikos Kottos}\email{Corresponding author: tkottos@wesleyan.edu}
\affiliation{Department of Physics, Wesleyan University, Middletown CT-06457, USA}

\begin{abstract}
We utilize caustic theory in ${\cal PT}-$symmetric lattices to design focusing and curved beam dynamics. We show that 
the gain and loss parameter in these systems provides an addition degree of freedom which allows for the design of the 
same caustics trajectories with different intensity distribution in the individual waveguides. Moreover we can create 
aberration-free focal points at any paraxial distance $z_f$, with anomalously large focal intensity.
\end{abstract}

\maketitle


{\it Introduction--} Diffraction management of beam propagation and the possibility of designing initial wavefronts which lead to abrupt energy focusing have attracted the attention of many researchers over the years \cite{ST07,SC07,SBDC07,GSWR11,KBNS12,DKSSA12,EC10,EC12}. Not only is the fundamental and mathematical physics side of this problem both charming and challenging for researchers, but also its applied side has attracted attention. Perhaps the most pronounced example comes from the field of medical lasers where one needs to have an abrupt beam focusing at a specific point, without affecting nearby tissues \cite{JLKHBM99}. Other applications include particle manipulation \cite{BMD08}, generation of self-bending plasma channels \cite{PKMSC09} and optical bullets \cite{CRCW10,ASPT10} etc.

\begin{figure}[htbp]
\centerline{\includegraphics[width=0.75\columnwidth]{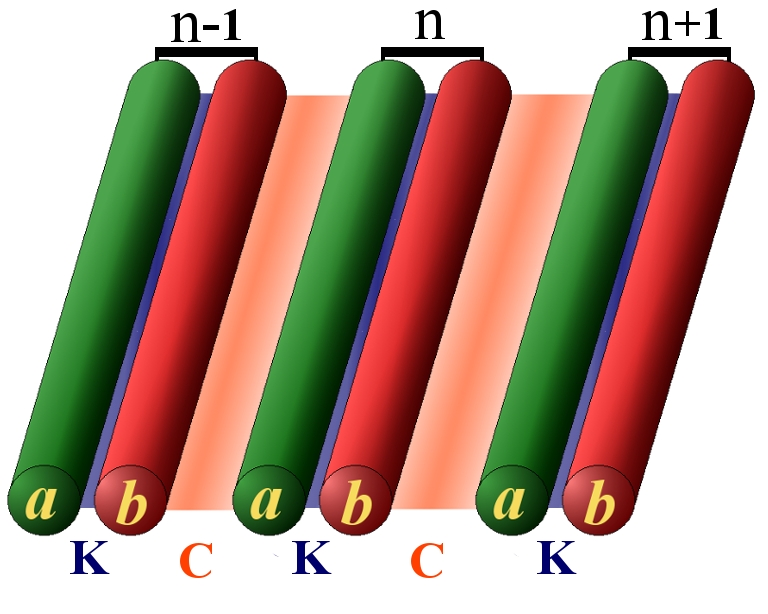}}
  \caption{A ${\cal PT}$-symmetric lattice consisting of dimers with local ${\cal PT}$-symmetry. Each dimer consists of two waveguides, one 
attenuating (type "a" waveguides, indicated with green) while the other one (type "b" waveguides, indicated with red) has equivalent amplification. 
The coupling between two waveguides of a specific dimer is $k$ while the coupling between different dimers is $c<k$.}
  \label{fig1}
\end{figure}

In this Letter we will investigate caustics trajectory dynamics and abrupt focusing in parity-time (${\cal PT}$)-symmetric discrete array settings, like the one shown in Fig. \ref{fig1}. These type of optical systems have been recently introduced \cite{MGCM08} and, during the last years, have gained a lot of attention due to the wealth of exotic properties that they posses \cite{MGCM08,RMGCSK10,a11,LREKCC11,FXFLOACS13,GSDMVASC12,RKGC10,BFBRCEK13,L10c,RLKKKV12,CJHYWJLWX14,POLMGLFNBY14,HMH,HMHCK14,L09,L09a,ZCFK10,S10,L10b,RCKVK12,RLKKKV12}. The main characteristic of these systems is that the optical potential (index of refraction) $\epsilon(x)=\epsilon_R(x) + i\gamma(x)$ is complex and satisfies the ${\cal PT}$ symmetric property $\epsilon(x)=\epsilon^*(-x)$. Since the optical potential is complex these systems are described by an effective non-Hermitian Hamiltonian with 
eigenvalues (propagation constants in optics language) which are real if the gain/loss parameter is below some critical value i.e. $\gamma\leq \gamma_{\cal PT}$. Above this value the eigenvalues become complex and the system is unstable. The transition point $\gamma=\gamma_{\cal PT}$ has the properties of an exceptional point (EP) singularity i.e. both eigenvalues and eigenvectors coalesce. This phase transition (coined ${\cal PT}$
-symmetric phase transition) leads to a number of interesting features: asymmetric transmission \cite{RKGC10,BFBRCEK13,CJHYWJLWX14,POLMGLFNBY14,RLKKKV12}, unidirectional invisibility \cite{LREKCC11,FXFLOACS13,GSDMVASC12}, novel lasing schemes \cite{LFLVEK14,HRZ,HMHCK14}, 
non-reciprocal Bloch oscillations \cite{L10b, L10c}, and reconfigurable Talbot effects \cite{RCKVK12}.

In this paper we will demonstrate the possibility of designing caustics and  identify the effects of non-hermiticity (i.e. gain/loss elements described by a complex index of refraction) in engineering beam trajectories and in the formation of abrupt focusing processes. Our approach will utilize curved trajectory dynamics and caustics design in discrete elements and thus extend the analysis for passive lattices \cite{EC12} to lattices with ${\cal PT}$
-symmetry.

{\it Model and theoretical analysis--} We consider an array of one-dimensional (1D) coupled waveguides. The array is composed of two types of waveguides: 
the first with gain (A) and the other with equivalent loss (B). We further assume that each waveguide is supporting only one propagating mode. The nearby 
waveguides are assumed to be coupled evanescently. The (A) and (B) waveguides are arranged in a way that they form $N$ coupled A-B dimers with intra-dimer coupling $k$ and 
inter-dimer coupling $c$. In the slowly varying envelope approximation the electric field amplitude $\Psi_n=(a_n,b_n)^T$ at the $n$-th 
dimer evolves (along the propagation direction $z$) according to the Schr\"odinger-like equation:
\begin{equation}
\label{dyndimer}
\begin{array}{lcr}
i \frac{da_n(z)}{dz}+\epsilon a_n(z) + k b_n(z) + c b_{n-1}(z) &=&0\\
i \frac{db_n(z)}{dz} +\epsilon^{\ast} b_n(z) + k a_n(z) + c a_{n+1}(z)&=& 0
\end{array}
\end{equation}
where the complex refractive index of the $n$-th waveguide is $\epsilon=\epsilon_0+i\gamma$. $\epsilon_0$ is the background index of refraction 
and $\gamma$ is the gain-loss parameter. Without any loss of generality, we will assume below that $\epsilon_0=0$. Equation (\ref{dyndimer}), which 
describes the field evolution, is invariant under a parity-time (${\cal PT}$) symmetric operation \cite{BFKS10,ZCFK10}. The parity-symmetry operator 
${\cal P}$ is defined as a spatial inversion (around an axis of symmetry of the array) while the time-reversal symmetry operator ${\cal T}$ is associated 
with a complex conjugation.

The propagation of the electric field along the paraxial direction $z$ is conveniently evaluated in the Fourier $q$ space where $a_n(z)=\frac{1}{2\pi}\int_{-\pi}^{\pi} dq {\tilde a}_q(z)\exp(inq)$ (similarly for $b_n$). Specifically, the translational invariance of the system allows us to decouple, for each value of $q$, the equations of motion into $2\times 2$ blocks: 
\begin{equation}
\label{dynfourier}
i\frac{d}{dz}\left(
\begin{array}{c}
{\tilde a}_q(z)\\
{\tilde b}_q(z)
\end{array}
\right)
= H_q\left(
\begin{array}{c}
{\tilde a}_q(z) \\
{\tilde b}_q(z)
\end{array}
\right)
;
H_q=\left(
\begin{array}{cc}
 -i\gamma & v_q\\
 v_q^{\ast} & i\gamma
\end{array}
\right)
\end{equation}
where $v_q=-(k+c\cdot e^{-iq})$. Equation (\ref{dynfourier}) can be solved analytically, thus allowing us to evaluate the wave packet evolution in the Fourier space. The wavepacket in the real space is then obtained by an inverse Fourier transform ${\tilde a}_q(z) = \sum_{m = -\infty}^{\infty} a_{m}(z) e^{-i q m}$ (similarly for ${\tilde b}_q(z)$).

The dispersion relation of the dimeric lattice is calculated from Eq. (\ref{dynfourier}) by substituting the stationary form $(a_n,b_n)^T=\exp(-i {\cal E}z)(A,B)^T$ \cite{ZCFK10}. We get:
\begin{equation}
\label{dispersion}
{\cal E}_{\pm}(q)= \pm\sqrt{(k-c)^2+4kc\cos^2(q/2) -\gamma^2}.
\end{equation}
It follows from Eq. (\ref{dispersion}) that for $\gamma <  \gamma_{\cal PT} = k-c$ the dispersion relation consists of two bands which are separated by a gap. In this parameter domain all the eigenvalues are real and the system is stable. The maximum gap size $ 2(k-c)$ occurs for $\gamma=0$. For larger values of $\gamma$ the gap becomes smaller until it disappears at $\gamma_{\cal PT}$. At $\gamma=\gamma_{\cal PT}$ the levels associated with $q= \pm\pi$ and their corresponding eigenvectors become degenerate resulting in an exceptional point (EP) singularity. For $\gamma>\gamma_{\cal PT}$ the spectrum becomes partially complex \cite{ZCFK10}. Bellow we assume that $k>c$.

Using the dispersion relation Eq.(\ref{dispersion}) we  evaluate the electric field amplitudes associated with the $n-$th dimer $\Psi_n=(a_n(z),b_n(z))^T$ at any distance $z$:
\begin{equation}
\label{eq4}
 \begin{pmatrix}
  a_{n}(z)\\
  b_{n}(z)\\
 \end{pmatrix} = \frac{1}{2 \pi} \sum_{m = -\infty}^{\infty} \begin{pmatrix}
  a_{m}(0)\\
  b_{m}(0)\\
 \end{pmatrix} \int_{-\pi}^{\pi}  e^{i q (n-m) - i {\cal E}_{\pm} z} dq.
\end{equation}
To proceed with our analysis we extend the integer index $m$, defining the dimer number, to a continuous variable $\xi$ which can be 
interpreted as the transverse spatial coordinate variable of the incident (i.e. $z=0$) field. Similarly, we express the integer index $n$ as 
a general continuous transverse spatial-coordinate x. This extension allow us to define the continuous smooth functions $a(\xi, 0)$ and 
$b(\xi, 0)$ with the property that $a(\left|\xi\right|\rightarrow \infty, 0)\rightarrow 0$ and similarly $b(\left|\xi\right|\rightarrow \infty, 
0)\rightarrow 0$. Next we express Eq.(\ref{eq4}) in the following integral form:
\begin{equation}
 \begin{pmatrix}
  a(x,z)\\
  b(x,z)\\
 \end{pmatrix} = \frac{1}{2 \pi} \int_{ -\infty}^{\infty}  \int_{-\pi}^{\pi}
\begin{pmatrix}
  \alpha_\xi\\
  \beta_\xi\\
 \end{pmatrix}  
e^{i(\phi_\xi + q (x-\xi) - {\cal E}_{\pm} z)} dq d\xi
\label{eq4b}
\end{equation}
where we have used the polar representation $a(\xi, 0) \equiv \alpha_\xi e^{i \phi_\xi}$ and $b(\xi, 0) \equiv\beta_\xi e^{i \phi_\xi}$. The subindex $\xi$ means that $\alpha,\beta$ and $\phi$ are functions of $\xi$. It turns out that the caustics formation is independent of the amplitudes $\alpha_{\xi};\beta_{\xi}$ (see Eqs. (\ref{eqcond},\ref{eqcond_c},\ref{q1}) below) and therefore we will assume below that $\alpha_{\xi}=\beta_{\xi}=1$. 

In order to enforce caustic beam dynamics, we impose a stationary phase (first and second order) condition on 
both integration variables $q$ and $\xi$ in Eq. (\ref{eq4b}): 
\begin{equation}
\label{eqcond}
\begin{array}{cc}
 \frac{\partial\Phi}{\partial\xi}=0\rightarrow q_\xi = \frac{\partial \phi_\xi}{\partial \xi}& (a)\\
  \frac{\partial\Phi}{\partial q}=0\rightarrow x = \xi - \frac{c k  z}{{\cal E}_{\pm}(q_\xi)}\sin q_\xi& (b)
\end{array}
\end{equation}
Along with the second order stationary condition
\begin{equation}
\label{eqcond_c}
\begin{array}{cc}
\frac{\partial^2\Phi}{\partial \xi^2}\frac{\partial^2\Phi}{\partial q^2} -(\frac{\partial}{\partial \xi}\frac{\partial\Phi}{ \partial q})^2 = 0
\rightarrow
 z = \frac{2\left(\frac{q_\xi{\cal E}_{\pm}^{3}}{c k q_\xi'}\right)}
{2 (c^{2} + k^{2} - \gamma^{2}) \cos q_\xi +c k (3 + \cos 2q_\xi) } 
\end{array}
\end{equation}
where $q_\xi '\equiv\frac{\partial q_\xi}{\partial\xi}$ and $\Phi = \phi_\xi + q (x-\xi) - {\cal E}_{\pm} z$. 
Equations (\ref{eqcond},\ref{eqcond_c}) define the coordinates $x$ and $z$ of the ray trajectory associated 
with the $\xi_{th}$ dimer. They allow us to design various types of caustics associated with the Eqs. (\ref{dyndimer}) 
of a beam propagating through the ${\cal PT}$-symmetric dimeric lattice of Fig. \ref{fig1}. 
Solving Eq. (\ref{eqcond}b) with respect to $q_\xi$ leads to the expression:
\begin{equation}
\label{q1}
q_\xi = \pm \cos^{-1}(-\eta^{2}\pm \sqrt{1 -\frac{(c^{2} + k^{2} -\gamma^{2})\eta^2}{ck} +\eta^4})
\end{equation}
where $\eta\equiv \frac{\xi-x}{\sqrt{c k}z}$. Further integration of Eq. (\ref{q1}) with respect to $\xi$ (see Eq. (\ref{eqcond}a)) 
results in a relationship between the lattice parameters, the ray trajectories, and the initial phases of the waveform $\phi_{\xi}$. 
This enables us to tailor the initial phases $\phi_{\xi}$ in order to achieve a desired caustics dynamics. The condition that the 
initial phases $\phi_{\xi}$ must be real imposes a constraint to $q_\xi$ that allows for the calculation of the maximum number 
of dimers $\xi\leq |\xi_{0}|$ that participate in the caustic dynamics.

We start with designing a focusing point at a predefined transverse coordinate position $x_f$ and at a paraxial distance $z_f$. 
Below we keep $z_f$ as a free parameter while, for the sake of presentation, we assert that $x_f=0$, i.e. it coincides with the 
origin of the transverse coordinate. The requirement for focusing at $x_f=0$ further simplifies Eq. (\ref{q1}), which can be 
used to evaluate the initial phase $\phi_\xi$ of the wavefronts. We get: 
\begin{equation}
\label{phi}
\phi_{\xi}^{\pm}= \pm\frac{\xi z_{f}}{|\xi|}  |{\cal E}_{\pm} (q_{f})| \pm \xi q_{f}+ z_{f} C_{0}
\end{equation}
\begin{equation}
\label{qf}
q_{\xi}^{f}=\cos^{-1}(-\eta_{f}^{2}+ \sqrt{1 -\frac{(c^{2} + k^{2} -\gamma^{2})\eta_{f}^2}{ck} +\eta_{f}^4})
\end{equation}
where $C_0= \sqrt{(c+k)^2-\gamma^2}$ and $\eta_{f}= \frac{\xi}{\sqrt{c k}z_{f}}$. The existence of a focal point for $z_{f}>0$ is achieved when the initial phases are $\phi_\xi^{+(-)}$ for $\xi<0 (\xi>0)$. The number of dimers $|\xi_0|$ that participate in the creation of the focal point can be calculated from Eq. (\ref{q1}). We get
\begin{equation}
\label{xi0}
\xi_{0} = -\frac{z_f \sqrt{c^2+k^2-\gamma ^2-C_0\sqrt{\gamma_{\cal PT}^{2}-\gamma^{2} }}}{\sqrt{2}}
\end{equation}

\begin{figure}[htbp]
\centerline{\includegraphics[width=0.85\columnwidth]{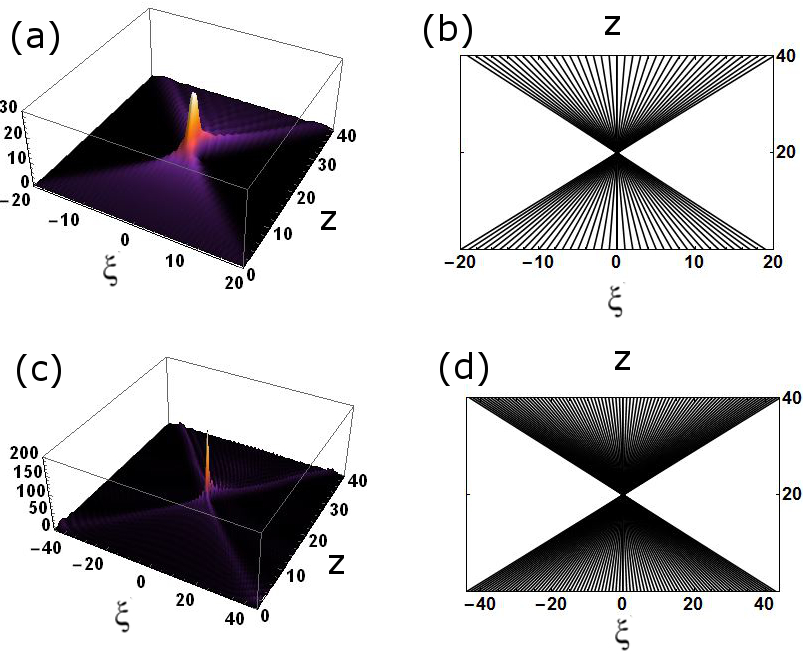}}
  \caption{Design of an aberration-free focal point at $z_f=20$ of an initial wavefront for a ${\cal PT}$-symmetric lattice with two different $\gamma$-values. 
In (a) and (c) the intensity profile of the evolving engineered beam is presented. In (b) and (d) we show the associated ray trajectories. Each individual ray 
corresponds to a dimer. The initial wavefront in (a) and (c) has the form $a(\xi, 0) = e^{i \phi_\xi}$ and $b(\xi, 0) = e^{i \phi_\xi}$ for $\xi_{0}\leq \xi \leq |\xi_{0}|$, 
where the initial phases $\phi_\xi$ are given from Eq.(\ref{phi}). The coupling constants are $k = 5$, and $c = 1$. In (a) and (b), $\gamma = 0$ and therefore we 
use $\left|\xi_0 \right|\approx 20$. In (c) and (d), $\gamma = 3.999 \approx \gamma_{\cal PT}$ and hence $\left|\xi_0\right| = 44$. In this case the field intensity 
at the focal point is an order of magnitude larger than in (a)}  
\label{fig5}
\end{figure}

Equations (\ref{phi},\ref{qf},\ref{xi0}) allow us to gain a better understanding of the role of the gain and loss parameter $\gamma$ 
in the formation of focusing points. Let us consider, for example, the two scenarios of a case with $\gamma=0$ and a ${\cal PT}$-
symmetric case with $\gamma =\gamma_{\cal PT}$. In the former case we find from Eq. (\ref{xi0}) that $|\xi_0|=c z_f$ dimer trajectories 
(and thus dimer waveguides) contribute for the construction of a focal point at a paraxial distance $z_f$ from the plane of preparation. 
In contrast when $\gamma \rightarrow \gamma_{\cal PT}$ we get $|\xi_0| = \sqrt{ck}z_f$ i.e. more waveguides are contributing to 
a focal point. In Fig. \ref{fig5} we show the evolution of an initial beam which is designed to create an ideal (i.e. aberration free) focal 
point for the two different values of $\gamma$. The phase engineered of the initial beam is dictated by the rules given by Eqs. (\ref{phi},
\ref{qf},\ref{xi0}) for each $\gamma$ value. We find that increasing the gain/loss parameter $\gamma$ leads to an increase of the number 
of waveguides $\xi_0$ which contribute to the creation of the focal point. At the same time the total beam intensity at the focal point 
increases dramatically. This growth is a consequence of the EP dynamics and has been predicted theoretically in \cite{ZCFK10} and 
observed experimentally in \cite{GSDMVASC12}. In the current framework this growth is manifested as a giant intensity growth at a 
pre-engineered focal point.

\begin{figure}[htbp]
\begin{center}
\includegraphics[width=0.85\columnwidth]{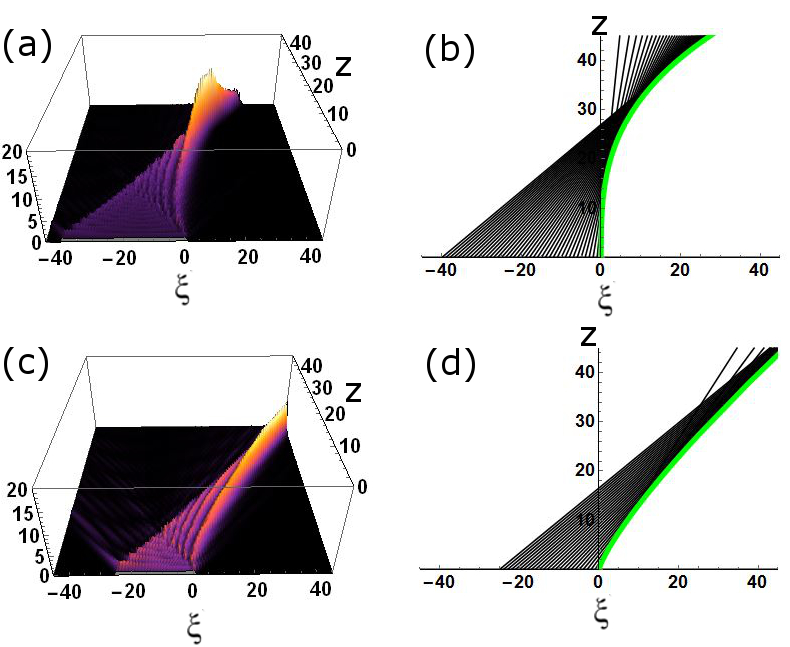}
\caption{Two representative examples of designed caustic propagation juxtaposed with the associated caustic trajectory of the form $x=\alpha z^{\delta}$. In both cases the lattice parameters are kept fixed $k=3$, $c=1$, and $\gamma =1.95$. The initial condition for the system is $a(\xi,0)=b(\xi,0)= \exp^{i \phi_{\xi}}$ for $\xi_{0} \leq \xi\leq 0$ and zero elsewhere. The difference between (a) and (c) is the caustic trajectory used to calculate the initial phase $\phi_{\xi}$. (b) is the associated ray trajectory for (a) and (d) is the associated ray trajectory for (c).  In (a) $\delta =3$, $\alpha =0.0003$ and consequently $\xi_{0} \approx -40$, while in (c) $\delta =5/4$, $\alpha =0.4$ and consequently $\xi_{0} \approx -25$. The green lines in (b) and (d) are the desired caustic trajectories.
\label{fig}}
\end{center}
\end{figure}

Next we investigate the management of curved caustic trajectories in ${\cal PT}$-symmetric lattices. We start with Eqs. (\ref{eqcond},\ref{eqcond_c},\ref{q1}). For theoretical simplicity the ensuing calculation will focus on designing caustics associated with an initial wave-front involving only dimers with index $\xi\leq 0$. We will assume a power law caustic trajectory $x =\alpha z^{\delta}$. Furthermore we relate the caustic propagation distance $z$, with the dimer index $\xi$ using Eq. (\ref{eqcond}b) which can be re-written in a more compact form as $\xi = x- z \frac{d x}{dz}$. One then uses these two relations to write the variables $x$ and $z$ as functions of the 
dimer index $\xi$, the power $\delta$ that defines the caustic trajectory, and the scaling constant $\alpha$. The resulting expressions are:
\begin{equation}
\label{x}
x = \frac{\xi}{(1-\delta)};\quad
z=(\frac{\xi}{\alpha(1-\delta)})^{1/\delta}
\end{equation} 
The above relations encode the information of the caustic trajectory into the ray trajectory coordinates for the $\xi$-th dimer. Substituting 
these expressions back into Eq. (\ref{q1}), enables us to calculate the initial phase distribution $\phi_{\xi}$ producing dynamics following the 
caustic trajectory defined by Eqs. (\ref{x}). Furthermore using Eq. (\ref{q1}) we estimate the number of dimers $\xi_0$ that are used to construct 
the propagating caustics. In Fig. \ref{fig}, we present two examples of different caustic trajectories for two different $\delta$ values. Our results 
confirm that power-law caustic trajectories can be supported in ${\cal PT}$-symmetric lattices. Obviously this agreement is valid up to some 
paraxial propagation distance $z$ and transverse coordinate variable $x$ which are given by Eq. (\ref{x}) for $\xi=\xi_0$.

\begin{figure}
\begin{center}
\includegraphics[width=0.75\columnwidth]{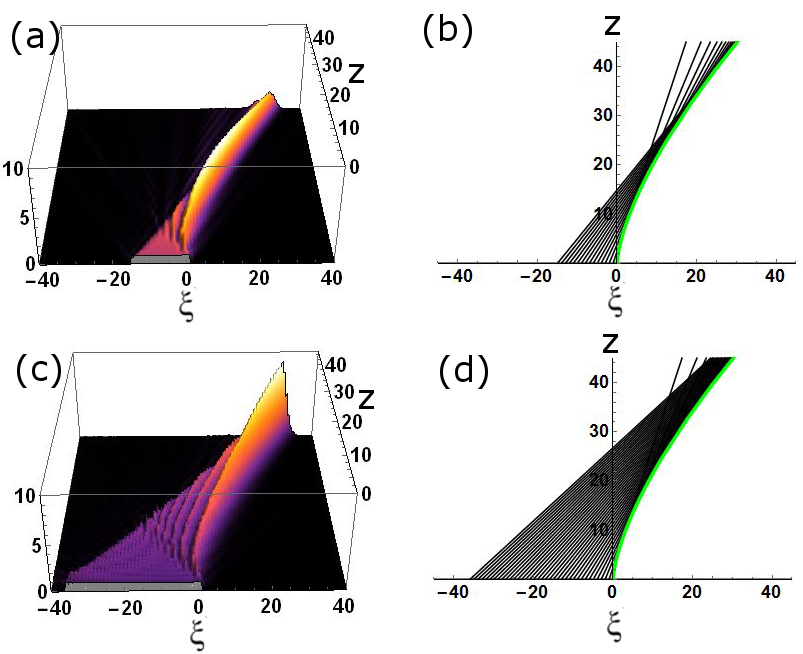}
\caption{Two examples of designed caustic propagation juxtaposed with the desired caustic trajectory of the form $x=\alpha z^{\delta}$, where $\alpha = 0.1$ and $\delta =3/2$, for (a) $\gamma =0.5$ and (c) $\gamma =1.8$. In both cases the lattice coupling parameters are kept fixed and equal to $k=3$, $c=1$. The initial wavefront is $a(\xi,0)=b(\xi,0)=\exp^{i \phi_{\xi}}$ for $\xi_{0} \leq \xi\leq 0$ and zero elsewhere.  In (a) $\xi_{0}=-15$ while in (c) $\xi_{0}=-36$. (b) is the associated ray trajectory for (a) while (d) is the associated trajectory for (c). In (a)-(b) $z_{\xi_{0}} = 45$ and $x_{\xi_{0}} = 30$ while in (c)-(d) $z_{\xi_{0}} = 80$ and $x_{\xi_{0}} = 72$. Furthermore, the point of maximal intensity along the caustic in (c) occurs approximately at the same coordinates as the end of the caustic in (a). The green lines in (b,d) are the desired caustic trajectories.
\label{fig2}}
\end{center}
\end{figure}

We can also study the effect of the gain/loss parameter $\gamma$ on a specific caustic trajectory. This is presented in Fig \ref{fig2} where we 
demonstrate the beam dynamics originated by the same caustic trajectory for two different values of gain/loss parameter. We see that increase 
of $\gamma$, while also adjusting the initial phase accordingly, augments the amplitude of the caustic wave without altering its trajectory. 

In conclusion we have investigated the possibility of implementing caustics design in ${\cal PT}$-symmetric lattices. We have demonstrated the design of power law caustic trajectories and the possibility of creating aberration-free focal points with increased focal power due to the presence of EP singularities. 

We acknowledge partial support from an AFOSR MURI grant FA9550-14-1-0037 and an NSF ECCS-1128571 grant.


\end{document}